\documentclass[useAMS,usenatbib]{mn2e}

\usepackage{epsfig}
\usepackage[totalwidth=480pt,totalheight=680pt]{geometry}

\newcommand{\bmain}{\begin{list}{${\bullet}$}{}}
\newcommand{\emain}{\end{list}}
\newcommand{\bminor}{\begin{list}{$\bf\triangleright$}{}}
\newcommand{\eminor}{\end{list}}
\newcommand{\Ms}{\mbox{$\,\mbox{M}_\odot$}}

\title[J0737-3039: Testing the Neutron Star Equation of State]
{The Double Pulsar J0737--3039: Testing the Neutron Star Equation of State}

\author[Ph.~Podsiadlowski et al.\ ]
{Ph.~Podsiadlowski$^{1}$\thanks{E-mail: podsi@astro.ox.ac.uk},
J.D.M. Dewi$^{1,2}$, P. Lesaffre$^{1,2}$, J.C. Miller$^{1,3}$,
\newauthor
W.G.~Newton$^{4}$, J.R. Stone$^{4,5,6}$
\smallskip
\\
\it
$^1$ Department of Astrophysics, University of Oxford, Oxford,
OX1 3RH\\
$^2$ Institute of Astronomy, Madingley Road, Cambridge, CB3 0HA\\
$^3$ SISSA and INFN, Via Beirut, 34014 Trieste, Italy\\
$^4$ Department of Physics, University of Oxford, Oxford, OX1 3PU\\
$^5$ Department of Chemistry and Biochemistry, University of Maryland, 
College Park, Maryland 20742, USA\\
$^6$ Physics Division, ORNL, Oak Ridge, TN 37831, USA
}

\date{\today}

\volume{000}

\pagerange{1--8} \pubyear{2005}

\begin{document}
\maketitle

\label{firstpage}

\begin{abstract}
The double pulsar J0737--3039 has become an important astrophysical
laboratory for testing fundamental physics.  Here we demonstrate that
the low measured mass of Pulsar B can be used to constrain the
equation of state of neutron star matter {\em under the assumption}
that it formed in an electron-capture supernova. We show that the
observed orbital parameters as well as the likely evolutionary history
of the system support such a hypothesis and discuss future refinements
that will improve the constraints this test may provide.
\end{abstract}

\begin{keywords}
binaries: close -- pulsars: general --
pulsars: individual: J0737--3039 -- stars: evolution -- stars: neutron
\end{keywords}

\section{Introduction}
The discovery of the first double pulsar, J0737--3039 (Burgay et al.\
2003; Lyne et al.\ 2004) consisting of two pulsars with spin periods
of 22.7\,ms (Pulsar A) and 2.77\,s (Pulsar B) in a 2.4\,hr orbit, has
opened up a new window for testing fundamental physics under extreme
conditions: not only will the system soon allow some of the best tests
of General Relativity (Kramer et al.\ 2005), but also it is the first
system where the physics of interacting magnetospheres from two
pulsars can be studied (e.g. Arons et al.\ 2005). The orbit is
decaying due to the loss of angular momentum by gravitational
radiation, and the two neutron stars are expected to merge in only
$\sim 85\,$Myr, a much shorter time than in any other double neutron
star system.  

%This has already led to an upward revision of the
%Galactic merger rate of DNSs (Burgay et al.\ 2003; Kalogera et al.\
%2004) with important implications for the direct detection of
%gravitational waves by present and future gravitational wave detectors
%and short-duration gamma-ray bursts, which may result from such
%mergers.

One of the other interesting characteristics of J0737--3039 is that
the mass of the second neutron star to be formed (Pulsar B) is the
lowest reliably measured mass for any neutron star to date: $1.249 \pm
0.001\Ms$ (Kramer et al.\ 2005). Such a low mass may be an indication
that the neutron star did not form in a standard iron-core-collapse
supernova but in an electron-capture supernova (Nomoto 1984;
Podsiadlowski et al.\ 2004)\footnote{Note, however, that it is not
entirely clear that electron-capture supernovae necessarily produce
the lowest remnant masses, since iron cores with masses as low as
$1.15\Ms$ may be unstable to collapse (see the discussion in Woosley,
Heger \& Weaver 2002).}. These occur for ONeMg white dwarfs when the
core density reaches a critical value at which electron captures
(e-captures) onto Mg (and subsequently Ne) start, causing a loss of
hydrostatic support in the core and triggering its collapse. One of
the key aspects of an e-capture supernova is that the collapse takes
places when the core density reaches a well-defined critical value
($\simeq 4.5\times 10^9\,$g\,cm$^{-3}$) which in turn occurs when the
ONeMg core has grown to a well-defined critical mass ($\simeq
1.37\Ms$; see \S~3). Therefore, if Pulsar B indeed formed in an
e-capture supernova, this would be the first instance for which the
masses of the pre-collapse core and the post-collapse neutron star
could both be determined, the former from a {\em theoretical\/}
estimate of the critical mass for an e-capture supernova, the latter
directly from the {\em observed\/} orbital parameters. Along with the
pre-collapse gravitational mass, the corresponding baryon number can
also be calculated. Since the loss of material during the formation of
the neutron star is expected to be extremely small in this scenario
(see \S~3), this is also a good approximation to the baryon number of
the neutron star. It is the purpose of this paper to demonstrate that
both the observed orbital parameters of the system (in particular the
low eccentricity) and the most likely evolutionary history of the
system favour the formation of Pulsar B in an e-capture supernova and
that comparison of its gravitational mass with the estimate obtained
for the baryon number enables useful constraints to be placed on the
neutron-star equation of state (EoS)\footnote{A different test of the
EoS of neutron-star matter can be derived from measuring the moment of
inertia of Pulsar A from the effects of spin-orbit coupling, as
proposed by Morrison et al.\ (2004).}.

In \S~2 we review the two most likely evolutionary channels that lead to
systems like the double pulsar and show how this discussion supports the
key assumption of Pulsar B having been formed in an e-capture supernova.
In \S~3 we provide a theoretical estimate for the pre-collapse core mass
(with an estimate of the uncertainties) and in \S~4 we demonstrate how the
properties of the system can be used to constrain the neutron star EoS.

\section{The Evolutionary History of J0737--3039}

\begin{figure*}
\psfig{file=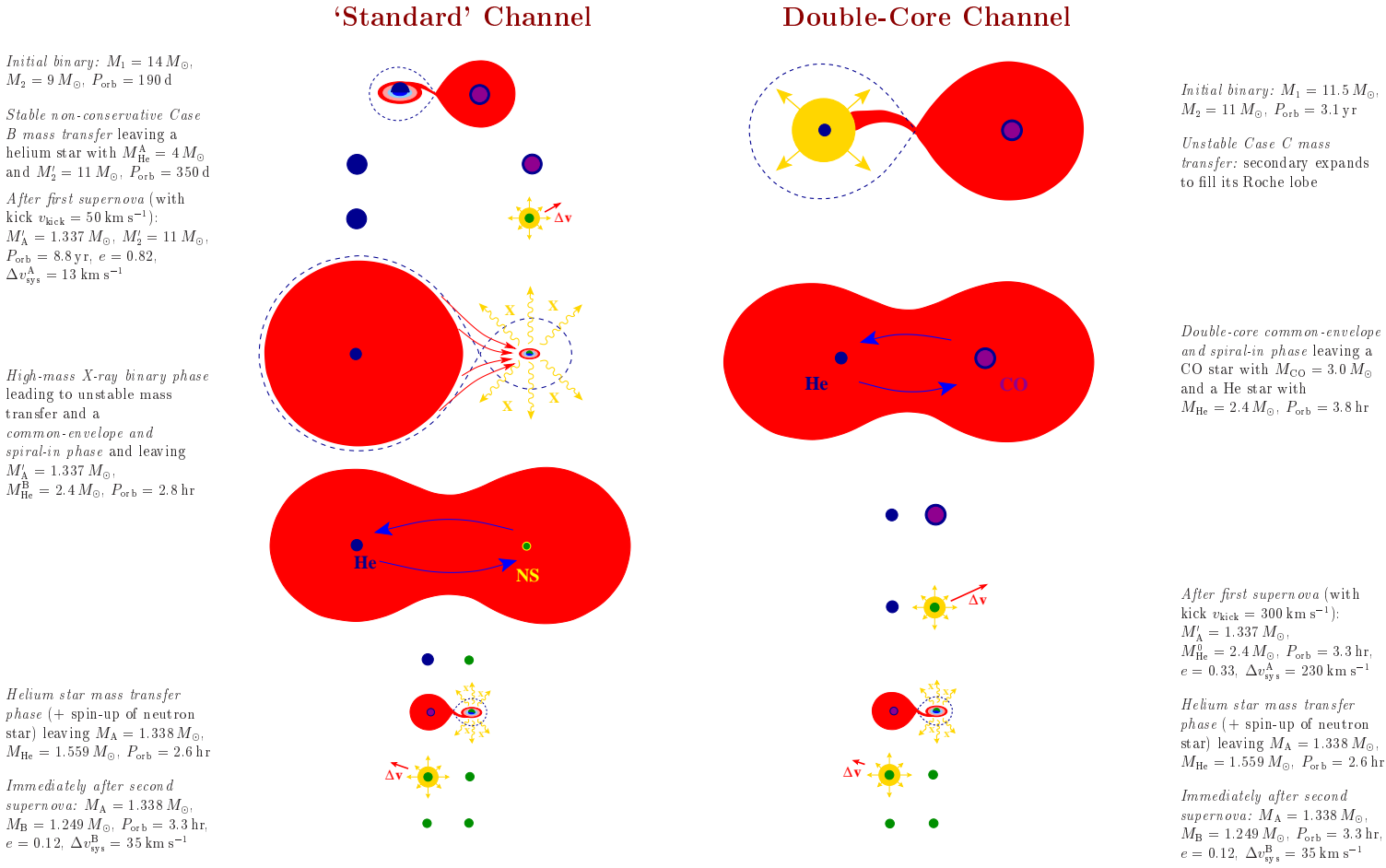,width=16cm,angle=-180}
\caption{Schematic diagrams illustrating the evolutionary history
producing systems like the double pulsar in the two main evolutionary
channels.  The final orbital parameters ($10^8\,$yr after the second
supernova) are identical to those of J0737--3039. The double-core
channel is expected to be less common than the standard channel
by a factor of 2 to 10. Note the similarity in the final evolutionary
stages, but the difference in kick-induced system velocities ($\sim
40$ and $\sim 230\,$km\,s$^{-1}$) in the two channels.}
\end{figure*}

There are two major evolutionary channels to form double neutron stars
like the double pulsar which we shall refer to in this paper as the
standard channel (e.g. Bhattacharya \& van den Heuvel 1991) and the
double-core channel (e.g. Brown 1995), respectively. In the standard
channel (see the left panel in Fig.~1), J0737--3039 originates from a
massive binary in which the more massive star transfers its envelope
to its companion star via stable Roche-lobe overflow (RLOF) before it
collapses in a supernova (SN) explosion to form the first neutron
star. If the SN explosion does not disrupt the system, the binary now
consists of a neutron star and a massive main-sequence star, and the
system will evolve into a high-mass X-ray binary (observationally it
will initially look like a Be X-ray binary). As the secondary evolves,
there will be a point at which it will fill its Roche lobe and start
to transfer matter to the neutron star. Because of the large mass
ratio of the system, this mass transfer is unstable and leads to a
common-envelope (CE) and spiral-in phase, in which the neutron star
spirals towards the centre of the massive companion inside the
companion's envelope. Provided that the CE phase does not lead to the
complete merger of the two components and that the CE is ejected, the
system evolves into a very close binary containing the helium core of
the secondary and the neutron star. Depending on the mass of the
helium star and the period of the system, another phase of mass
transfer may occur, where the neutron star is spun up and becomes a
fast `recycled' pulsar. Eventually the helium star collapses to form
the second neutron star in the system.

In the double-core channel (see the right panel in Fig.~1), the binary
components are very close in mass initially (within 5\,--\,10\,\%),
and the orbit is relatively wide, so that the primary only fills its
Roche lobe after it has completed helium core burning (so-called Case
C mass transfer) and has developed a CO core. At this stage, the
secondary has already completed its hydrogen core burning phase and
has evolved off the main sequence. Because of the high mass transfer
rate, the accreting star expands to fill and ultimately overflow its
Roche lobe, and the system is again expected to enter into a CE phase;
but, in this case, the CE is formed from the {\em combined} envelopes
of both stars. Inside the CE, there are the cores of the two stars,
the more evolved one with a CO core and the less evolved He core of
the secondary. The cores spiral-in inside the joint envelope until the
envelope is ejected, leaving a very close binary consisting of the two
evolved cores. The CO core soon collapses to form the first neutron
star, leaving a binary consisting of a neutron star and a helium
star. The further evolution is almost identical to the evolution in
the standard scenario.

Because of the constraints on the initial mass ratio and the orbital
separation, the double-core channel is expected to have a lower
occurrence rate than the standard channel (by a factor of 2 to 10;
Dewi, Podsiadlowski \& Sena 2005; in preparation)\footnote{However, if
the spiral-in of a neutron star in a massive envelope always leads to
hypercritical accretion onto the neutron star and its conversion into
a black hole, as argued, e.g., by Brown (1995), the double-core
channel would be the only one of these two channels that could produce
double neutron star systems (also see Chevalier 1993).}.

The formation of J0737--3039 via the standard channel has been studied
by Dewi \& van den Heuvel (2004) and Willems \& Kalogera (2004); both
studies concluded that this system must have originated from a close
helium star--neutron star (HeS-NS) binary where the system underwent
mass transfer during the helium-star phase, spinning up the
first-born neutron star in the process. Because the final stages of
evolution in the standard and the double-core channel are essentially
the same, i.e. the HeS-NS phase, the following discussion, which
assumes the standard channel, also applies to the double-core channel.

For different reasons, Dewi \& van den Heuvel (2004) and Willems \&
Kalogera (2004) found a similar pre-SN mass of the helium star
progenitor of J0737--3039. Willems \& Kalogera (2004) took the
threshold helium-star mass for the formation of a neutron star to be
2.1~$\mathrm{M_{\odot}}$ (Habets 1986) as the lower limit. However,
one should note that after the mass-transfer phase, the immediate
pre-supernova mass can be as low as the Chandrasekhar mass ($\sim
1.4\Ms$). Dewi \& van den Heuvel (2004) used 2.3~$\mathrm{M_{\odot}}$
as the minimum possible helium-star mass at the time of the explosion,
based on the assumption that lower-mass helium stars experience a
further CE phase at the end of their evolution (Dewi et al. 2002; Dewi
\& Pols 2003). This lower limit on the pre-SN mass then required a
minimum kick velocity of 70~$\mathrm{km \, s^{-1}}$, since the 
supernova mass loss on its own would produce a much larger 
post-supernova eccentricity than is consistent with the present
orbital parameters of J0737--3039.

However, the assumption of the occurrence of another CE phase in a
lower-mass helium star is still an open question. A recent population
synthesis study of the formation of double neutron stars (Dewi, Pols
\& van den Heuvel, 2005, in preparation) suggests that, particularly
to explain the formation of J0737--3039, it is more likely that this
CE phase does not occur.
%
%This
%is because their simulations show that the most probable progenitors
%of the double pulsar are helium stars with initial mass less than
%$\sim$ 3.2~$\mathrm{M_{\odot}}$. After RLOF, the pre-SN masses are
%less than $\sim$ 1.8~$\mathrm{M_{\odot}}$. As found by Dewi \& Pols
% (2003), at the end of the evolution of helium stars less than $\sim$
%3.2~$\mathrm{M_{\odot}}$ the ONeMg core masses are higher than the
%threshold mass for neon ignition, 1.37~$\mathrm{M_{\odot}}$ (Nomoto
%1984). However, the shell with maximum temperature in the core does
%not move inward to the centre. Hence, neon will be ignited off-centre,
%and it is likely that helium stars in this range of mass do not
%undergo core collapse due to photodisintegration, but due to electron
%capture.
% {\bf can Pierre's work support this argument?}
%
In this case, the final pre-supernova mass can be much less than
2.3~$\mathrm{M_{\odot}}$, indeed it can be as low as $\simeq 1.4\Ms$,
and no supernova kick is required to compensate for the mass loss; in
particular it allows for the possibility that the supernova was
symmetric (Dewi \& van den Heuvel 2005), strongly favouring an
e-capture SN for the second supernova.

\subsection*{An electron-capture supernova to form Pulsar B?}

The eccentricity of the double pulsar binary ($e\simeq 0.088$ at
present, most likely 0.11\,--\,0.12 immediately after the second
supernova; e.g. Burgay et al.\ 2003) is surprisingly low, much lower
than one would expect if the system received a large supernova kick in
the second supernova that formed Pulsar B.  Indeed, such a low
eccentricity can be most easily explained by a symmetric second
supernova in which a moderate amount of mass is expelled from the
system. In this case the eccentricity is given by $e = \Delta
M/(M_{\rm A}+M_{\rm B})$, where $\Delta M$ is the mass lost in the
supernova and $M_{\rm A}$ and $M_{\rm B}$ are the present masses of
Pulsars A and B, respectively. Taking $M_{\rm A} = 1.338\Ms$ and
$M_{\rm B} = 1.249$ (Kramer et al.\ 2005) and assuming a post-SN
eccentricity $e_0=0.12$ then yields a pre-SN mass of the helium star
of $\simeq 1.56\Ms$. Such low-mass pre-SN helium stars typically form
from HeS-NS binaries with initial helium stars of less than $\sim
3\Ms$ (see Dewi et al.\ 2002; Ivanova et al.\ 2003). This includes the
mass range where helium stars are expected to end their evolution in
an e-capture supernova (Nomoto 1984).  This is also consistent with
the speculation by Podsiadlowski et al.\ (2004) that e-capture
supernovae may not produce large supernova kicks since the explosion
may proceed on a timescale that is much shorter than the timescale on
which the instabilities that produce large kicks can develop; this
suggestion has received some theoretical support from recent
core-collapse calculations (Scheck et al.\ 2004; H.-Th.\ Janka 2005
[private communication]). Since in any {\em simple} accretion model
one would expect that the spin of Pulsar A would become aligned with
the orbital momentum axis and since this orientation is not affected
by a symmetric supernova, this may make the testable prediction that
the post-SN misalignment angle between the spin of pulsar A and the
orbital axis should be relatively small. This may indeed account for
the surprising stability of the pulse shape of Pulsar A (Manchester et
al.\ 2005).

\subsection*{The second supernova kick and the space velocity of
J0737--3039}

A low supernova kick velocity, as suggested by the low eccentricity,
would typically also imply that the binary system should only have
received a relatively small additional kick in the second supernova
and that the system space velocity relative to the local standard of
rest would not be much affected by it. Indeed, the space velocity of
J0737--3039 may provide an important constraint on its evolutionary
history. Unfortunately, the situation regarding the value of this
quantity is at present somewhat confused. Using interstellar
scintillation to measure the transverse space velocity of the system,
Ransom et al.\ (2004) determined a large system velocity of at least
140\,km\,s$^{-1}$.  Subsequently, Coles et al.\ (2005) showed that
the dispersion across the field was highly variable reducing the
estimate of the scintillation velocity to a value as low as
66\,km\,s$^{-1}$ and possibly even lower, since this value does not
account for the motion of the Earth. More recently, Kramer et al.\
(2005) argued that the present limits on the proper motion of the
system suggest a low transverse velocity of less than
30\,km\,s$^{-1}$, which would imply that it is statistically unlikely
that the system received a large kick in the second supernova. On the
basis of the original high estimate of the space velocity, Ransom et
al.\ (2004) and Willems et al.\ (2004) concluded that the system
should have received a large kick in the second supernova. However,
they only considered the standard scenario in which the system is
still fairly wide at the time of the first supernova; this implies
that it cannot receive a large kick in the first supernova and
remain bound (see Fig.~1). In contrast, in the double-core scenario
where the system is already very tight at the time of the first
supernova, the system is expected to receive a large systemic kick
from the first supernova (of order 150\,--\,400\,km\,s$^{-1}$: see
Fig.~1 and Dewi et al.\ 2005, in preparation) and hence no additional
kick from the second supernova would be required. As far as testing
the EoS is concerned, it is not important whether the system velocity
is low or high, since both can be understood within the framework of
either of the two channels and a small second kick.  In particular, we
note how similar the final phases are in the two channels. A
resolution of the issue of the system's space velocity could, however,
provide a powerful discriminant between the standard channel and the
double-core channel.

\section{Electron-Capture Supernovae}

An electron-capture supernova occurs when the central density of an
ONeMg core reaches the threshold value $\rho_{\rm th}$ for electron
captures on {$^{24}$Mg}. This decreases the electron pressure and the
electron fraction $Y_e$, lowering the Chandrasekhar mass and
triggering the collapse of the core (Miyaji et al.\ 1980).

In order to estimate the uncertainties in the critical mass of the
collapsing core, we performed a series of stellar structure
calculations assuming non-rotating cores in hydrostatic equilibrium with a
prescribed central density, homogeneous composition and a specified
thermal profile. Since the heat released by the electron captures
gives rise to a convective core, we adopted an isentropic thermal
profile.

As our reference model we adopted the composition given by Guti\'errez
et al.\ (1996) with X($^{16}$O)=0.72, X($^{20}$Ne)=0.25,
X($^{24}$Mg)=0.03, central density $\rho_{\rm th}=4.5\times
10^9$\,g\,cm$^{-3}$ and a range of central temperatures from $10^7$ to
$10^9$~K.  We used the equation of state of Pols et al.\ (1995),
assuming full ionization (i.e. we discard the ionization pressure
term). Note that neglecting the Coulomb corrections in the equation of
state would increase the total mass by $3.79\times 10^{-2}$M$_\odot$
for our reference composition.

We integrated the general relativistic (GR) equations of hydrostatic
equilibrium out from the centre
\begin{equation}
\frac{{\rm d}P}{{\rm d} r}=-\frac{Gm\rho}{r^2}\left(1+\frac{P}{\rho c^2}\right)\,
	\left(1+\frac{4\pi r^3 P}{m c^2}\right)\,
    \left(1-\frac{2 G m}{r c^2}\right)^{-1},
\end{equation}
where
\begin{equation}
\frac{{\rm d}m}{{\rm d} r}=4 \pi \rho r^2,
\end{equation} and
\begin{equation}
\frac{{\rm d} A}{{\rm d} r}=4 \pi n_b\,r^2\,
\left(1-\frac{2 G m}{r c^2}\right)^{-\frac12}.
\end{equation}
Here $m$ and $A$ are the gravitational mass and baryon number enclosed
within a sphere of radius $r$; $\rho$ is the density, including
contributions from both the rest mass and the thermal energy; $n_b$ is
the baryon number density. Rather than talking in terms of the baryon
number $A$, which is a rather abstract quantity, it is convenient to
convert this into a mass by multiplying by the atomic mass unit
($931.50\,$Mev/$c^2$), and we refer to this quantity as the {\em baryonic
mass}. If we had ignored the general relativistic corrections, our
estimate of the baryonic mass would have been increased by $1.30\times
10^{-2}$M$_\odot$ ($\sim 1\%$).  This difference is larger than might
have been expected; the reason, however, is analogous to
that for the large difference between the Newtonian and GR
values for the maximum mass of a neutron star (Oppenheimer \& 
Volkoff 1939).

% but is a consequence of the fact that all
%the GR corrections act to lower the critical mass\footnote{This is
%similar to the large effect GR corrections have for the maximum mass
%of a star supported by neutron degeneracy pressure, i.e. the
%Chandrasekhar mass for neutron stars, lowering the maximum mass from
%5.8\Ms\ to 0.7\Ms\ (Oppenheimer \& Volkoff 1939).}.

To check the validity of our procedure, we computed a model without the
GR corrections to reproduce the case considered by Miyaji et al.\ (1987) 
and Nomoto (1987), where they used a composition $X(^{16}$O$)=0.12$,
$X(^{20}$Ne$)=0.76$, $X(^{24}$Mg$)=0.12$, central density $3.98\times
10^9$\,g\,cm$^{-3}$ and central temperature $\log(T_c)=8.61$.  We find
a total mass $m=1.3754$\,M$_\odot$, which is very close to their
published value of 1.375\,M$_\odot$.

This composition is, however, no longer considered appropriate since
the reaction rate {$^{12}$C}($\alpha,\gamma$){${^{16}}$O} has been
revised upwards (Dominguez, Tornambe \& Isern 1993). This leads to a
lower Ne and Mg abundance and a higher O abundance at the end of C
burning than computed with the Fowler et al.\ (1975) rates (as was
done in Miyaji et al.\ 1987). To estimate the uncertainties introduced
by the composition, we investigated several different
compositions found in the literature (Dominguez et al.\ 1993;
Guti\'errez et al.\ 1996; Ritossa, Garc\'\i a-Berro \& Iben 1996;
Gil-Pons \& Garc\'\i a-Berro 2001) after the revision of the
{$^{12}$C}($\alpha,\gamma$){${^{16}}$O rate. The composition is
important in determining the Coulomb parameter and $Y_e$, and hence
the magnitude of the Coulomb corrections.  We use our reference model
to give a lower bound on the critical mass (solid curve in Fig~2) and
the composition $X(^{16}$O$)=0.56$, $X(^{20}$Ne$)=0.29$,
$X(^{24}$Mg$)=0.06$, $X(^{23}$Na$)=0.07$ and $X(^{12}$C$)=0.01$, which
mimics the composition of Gil-Pons \& Garc\'\i a-Berro (2001) and has
the lowest Coulomb correction, to give an upper bound to the critical
mass (dashed curve in Fig~2).

Finally, we changed the threshold density $\rho_{\rm th}$ from
$4.5\times 10^9$ to $4\times 10^9$\,g\,cm$^{-3}$ in order to estimate the
effect of the shift in the critical density due to the Coulomb
corrections (the latter is the appropriate density without any Coulomb
corrections; see Guti\'errez et al.\ 1996 for a detailed discussion).
This change decreases the critical mass by $2\times 10^{-3}$M$_\odot$
(dotted and dot-dashed lines in Fig~2).

In the previous studies, the central temperature after the onset of
electron captures ranged from $\log T_c =8.5$ up to $\log T_c =8.65$.
For this temperature range, Figure~2 then implies that the baryonic
mass of the pre-collapse core should lie between 1.366 and 1.375\Ms\
(the thin dotted curves) taking into account the uncertainties in
temperature and composition. On the assumption that the loss of
material during formation of the neutron star is negligible, this then
gives the predicted range for the baryonic mass of the neutron star
(which we refer to as $M_0$).

\subsection*{Caveats}

At this point, several caveats should be made about other effects
which may systematically affect this estimate. First, we take the
baryonic mass of the pre-collapse ONeMg core to be the same as that of
the neutron star, neglecting any loss of material during the formation
of the neutron star.  In practice, some material may be ejected in the
supernova in a neutrino-driven wind (Qian \& Woosley 1996).  However,
because of the steep density gradient at the edge of the ONeMg core,
we expect this mass loss to be small, probably less than a few times
$10^{-3}\Ms$, although it could potentially be as large as
$10^{-2}\Ms$ and the amount of mass loss itself depends on the EoS
(Janka 2005, private communication). Second, an e-capture supernova
may occur between carbon shell flashes when the central density
increases significantly (although we note that this point is not yet
fully resolved). Because of the discrete nature of the carbon flashes,
this may introduce a natural variation in the critical mass from star
to star because of variations of the thermal profile in the outer
ONeMg core. In this context we note that the assumption of an
isentropic profile is unlikely to be correct in the outer parts of the
ONeMg core.

% and that the GR corrections can be
%quite sensitive to this.

It is clear that these effects could significantly change our estimate
for the baryonic mass of the neutron star (probably decreasing it) and
therefore our present estimates should only be considered as
preliminary. However, with the expected progress in simulating
e-capture supernovae and calculating the evolution of their
progenitors (with improved e-capture rates, a richer nuclear network,
inclusion of accurate Coulomb corrections and using GR), we estimate
that ultimately one should be able to pin-point the mass of the
collapsing core to within  $\sim 2\times 10^{-3}$M$_\odot$.

 \begin{figure}
 \psfig{file=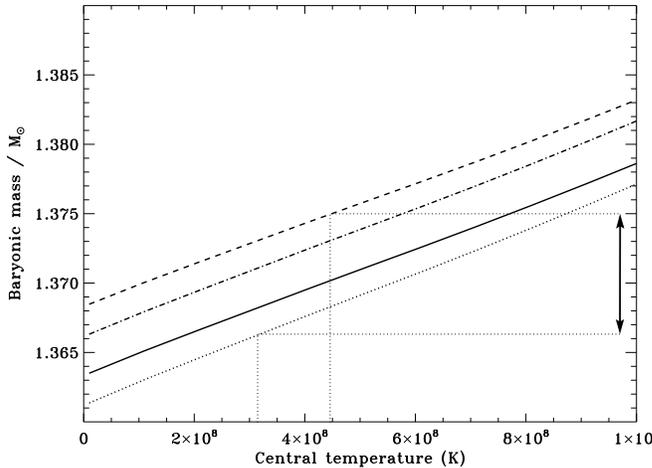,angle=-90,width=9cm}
\label{fig}

\caption{Baryonic mass versus central temperature for various ONeMg
cores in hydrostatic equilibrium. Solid curve: reference model with
$\rho_{\rm th} = 4.5\times 10^9\,$g\,cm$^{-3}$ with O and Ne mass
fractions $X(^{16}{\rm O}) = 0.72$ and $X(^{20}{\rm Ne})=0.25$
(minimum mass for the variation in composition). Dashed curve: similar
but with $X(^{16}{\rm O}) = 0.56$ and $X(^{20}{\rm Ne})=0.29$ (maximum
mass for the variation in composition).  Dotted curve: $\rho_{\rm th}$
decreased to $4\times 10^9\,$g\,cm$^{-3}$ with the composition of
the reference model.  Dot-dashed curve: same model as for the dashed
curve but with decreased $\rho_{\rm th}$. The thin dotted lines give
the most likely range of parameter values at the time of core collapse.}
\end{figure}

\section{Constraints on the Equation of State}

On the hypothesis that the scenario presented in the previous sections is
correct, we can use the information about the gravitational and baryonic
masses of Pulsar B to place constraints on the EoS of neutron-star matter.
These turn out to be quite interesting.

For any given EoS for neutron-star matter, one can calculate the
relation between the gravitational mass and the baryonic mass (bearing
in mind that the rotation speed is so low that taking the object to be
spherical and non-rotating is an excellent approximation). The present
gravitational mass (known from observations) and the baryonic mass
(known from the stellar evolution calculations) then define an error
box through which the relations calculated from the equations of state
would need to pass in order to be consistent. In this section, we
discuss how the constraints obtained in this way turn out. The
observed gravitational mass ($M_G = 1.249 \pm 0.001 \Ms$) and the
calculated baryonic mass ($M_0$ in the range $1.366 - 1.375 \Ms$)
specify the boundaries of the error box.

Results obtained by integrating the GR equations of hydrostatic
equilibrium (Eqs~1\,--\,3) for a range of EoSs are shown in the four
panels of Figure 3, each corresponding to a particular class of
equations of state (we use a modified form of the classification in
the paper by Morrison et al.\ 2004). All of these EoSs give the
maximum gravitational mass $M_{\rm max}$ for a non-rotating neutron
star as being above $1.5\Ms$, in line with observations. Class I EoSs
(top left-hand panel) come from non-relativistic many-body
calculations with ``realistic'' potentials: APR98 (Akmal,
Pandharipande \& Ravenhall 1998), WFF88 (Wiringa, Fiks \& Fabrocini
1988) and FPS (Lorenz, Ravenhall \& Pethick 1993) include only
nucleonic degrees of freedom, BJ74 (Bethe \& Johnson 1974) and MOSZ74
(Mozskowski 1974) include also hyperonic components at the higher
densities. Class II EoSs (top right-hand panel) use relativistic
mean-field (or effective-field) approximations including hyperonic
degrees of freedom: GLE210, GLE240 and GLE300 (Glendenning 2000), and
HOF01 (Hofmann et al.\ 2001) are shown. We also include here one EoS
representing a hybrid stellar model (nucleons + quarks): GLENHYB
(Glendenning 2000). Class III EoSs (bottom left-hand panel) use
non-relativistic phenomenological potentials of the Skyrme type (see
Stone et al.\ 2003 and references therein). Class IV EoSs (lower
right-hand panel) are for other phenomenological non-relativistic
potentials: BPAL21 and BPAL31 (Prakash et al.\ 1997) have only
nucleonic degrees of freedom, while BAL97 (Balberg \& Gal 1997)
includes hyperons at high density. All of these high-density EoSs were
joined onto the Baym-Bethe-Pethick EoS (Baym, Bethe \& Pethick 1971)
at a density of $\sim 1.4\times10^{14}$ g$\,$cm$^{-3}$ ($0.08 - 0.09
\,$fm$^{-3}$) and, in turn, this was joined onto the
Baym-Pethick-Sutherland EoS (Baym, Pethick \& Sutherland 1971) at $4.2
\times 10^{11}$ g$\,$cm$^{-3}$ ($2.5 \times 10^{-4}\,$fm$^{-3}$). By
doing this, the inner and outer crust of the neutron star were treated
in the same way for all of the EoSs, and so all of the differences
seen result from differences in the treatment of the high-density
matter. (Strange star models have not been included in Fig~3; they
cover a very wide range and give curves passing both above and below
the error box as well as curves passing through it.)

The most clear-cut result is that none of the class II models tested
in this work give predictions in line with our constraint. For the
other classes (I, III and IV), the situation is less clear and depends
on the particular properties of each individual EoS. For the
phenomenological Skyrme potentials, the EoSs give a wide range of
predictions within the region delimited by SkI1 and BSk8. Those
parametrisations giving $M_G/M_0$ curves passing through the error box
give $M_{\rm max}$ between $1.6 - 1.9\Ms$. All of those for which
$M_{\rm max} > 1.9\Ms$ give curves passing above the error box while
those for which $M_{\rm max} < 1.6\Ms$ give curves passing below
it. With reference to the discussion in Stone et al.\ (2003) and using
the notation of that paper, we note that all of the Skyrme EoSs
passing through the error box are type II parametrisations whereas all
of those coming from type I parametrisations pass above it.

Apart from the comments made above, there is no simple general
interpretation of the implications of our constraint for the physics
behind the particular EoSs. It is important to recognise that our
constraint (assuming that our overall scenario is correct) represents
a necessary but not sufficient condition for choosing a suitable
EoS for neutron star models.  Additional observational information is
needed, in particular concerning the neutron star maximum mass which,
in combination with the present constraint, would give a more
definitive criterion for the choice of physical model for the
EoS. Also, the influence of different treatments for the inner and
outer crust needs to be investigated before any final conclusion is
drawn.

\begin{figure*}
\label{fig}
\centerline{\psfig{file=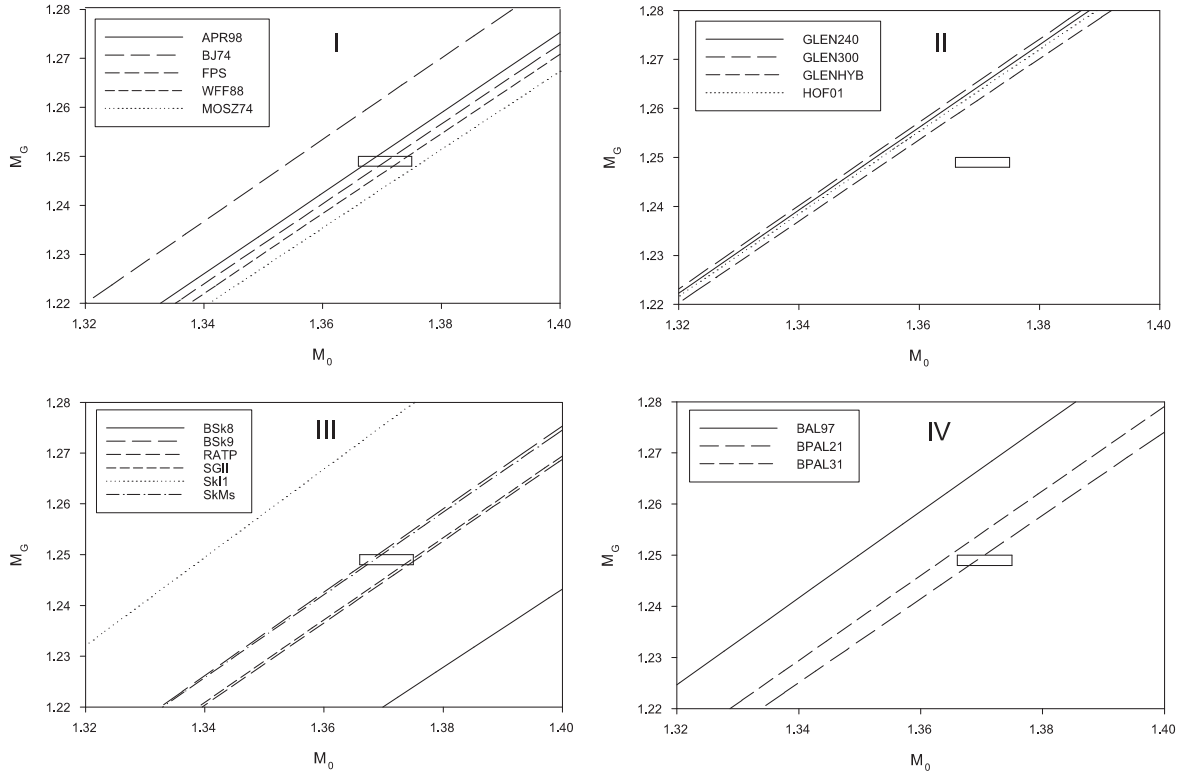,width=16cm}}
\caption{Relation between the gravitational mass M$_G$ 
of neutron star models and their baryonic mass M$_0$, measured in 
units of the solar mass \Ms, for various equations of state.
The constraint derived in this paper is marked by a rectangle. For
more explanation, see text.}

\end{figure*}

\section{Conclusions}

In this paper, we have demonstrated that the measured gravitational
mass of Pulsar B in the double pulsar J0737--3039 can be used to give
a new test of the neutron-star EoS if one makes the all-important
assumption that Pulsar B was formed in an electron-capture
supernova. In this case, its baryonic mass can be estimated
theoretically and comparison between this and its gravitational mass
can then be used to constrain the EoS. We have re-constructed the
possible evolutionary histories for J0737--3039 in the main formation
channels to support the hypothesis of Pulsar B having originated in an
electron-capture supernova and have discussed possible tests of this
hypothesis. Future refinements, both of the stellar evolution models
(to better pin down the critical pre-collapse mass) and of
electron-capture collapse models (to quantify the possible mass loss),
should lead to a more stringent constraint which, when combined with
other astrophysical EoS constraints, can provide new insight into the
physics of neutron-star matter.

\section*{Acknowledgements}
We thank H.-Th.\ Janka and M.\ Kramer for very useful discussions and
sharing the results of some of their unpublished work and C.M.Keil for
supplying numerical data for EOS HOF01 in Figure~3.  This work was in
part supported by a European Research \& Training Network on Type Ia
Supernovae (HPRN-CT-20002-00303, PhP, PL), a Talent Fellowship (JDMD)
from the Netherlands Organization for Scientific Research (NWO), EPSRC
Grant 02300018 (WN), an Advanced Computing Grant from US DOE
Scientific Discovery (JRS) and US DOE grant DE-FG02-94ER40834 (JRS).

\label{lastpage}

\end{document}